\documentclass[floatfix,prstab,showpacs,  preprint]{revtex4}
\usepackage{lineno}
\usepackage{natbib}
\usepackage{graphicx}
\usepackage{hyperref}
\usepackage{amssymb}
\usepackage{amsmath}
\usepackage{dcolumn}
\usepackage{multirow}

\begin{document}

\title{Photocathode Behavior During High Current Running in the Cornell ERL Photoinjector}
\author{Luca Cultrera}\email[Corresponding Author: ]{lc572@cornell.edu}
\author{Jared Maxson} \email[Corresponding Author: ]{jmm586@cornell.edu}
\author{Ivan Bazarov} \email{ib38@cornell.edu}
\author{Sergey Belomestnykh} \email[Now at Brookhaven National Lab, P.O. Box 5000, Upton NY 11973: ]{sbelomestnykh@bnl.gov}
\author{John Dobbins}
\author{Bruce Dunham}
\author{Siddharth Karkare}
\author{Roger Kaplan}
\author{Vaclav Kostroun}
\author{Yulin Li}
\author{Xianghong Liu}
\author{Florian L\"ohl}
\author{Karl Smolenski}
\author{Zhi Zhao}
\author{David Rice}
\author{Peter Quigley}
\author{Maury Tigner}
\author{Vadim Veshcherevich}
\author{Kenneth Finkelstein}
\author{Darren Dale}
\author{Benjamin Pichler}

\affiliation{\\
  Cornell Laboratory for Accelerator-Based Sciences and Education,
  Cornell University, Ithaca, New York 14853
}


\begin{abstract}
The Cornell University Energy Recovery Linac (ERL) photoinjector has recently demonstrated operation at  20 mA for  approximately 8 hours, utilizing a multialkali photocathode deposited on a Si substrate. We describe the recipe for photocathode deposition, and will detail the parameters of the run. Post-run analysis of the photocathode indicates the presence of significant damage to the substrate, perhaps due to ion back-bombardment from the residual beamline gas. While the exact cause of the substrate damage remains unknown, we describe multiple surface characterization techniques (X-ray fluorescence spectroscopy, X-ray diffraction, atomic force and scanning electron microscopy) used to study the interesting morphological and crystallographic features of the photocathode surface after its use for high current beam production. Finally, we present a simple model of crystal damage due to ion back-bombardment, which agrees qualitatively with the distribution of damage on the substrate surface.
\end{abstract}
\maketitle

\section{Introduction}
X-ray light sources based on Energy Recovery Linacs (ERLs) and Free Electron Lasers (FELs) rely primarily on photoemission to generate a high brightness photon beam. The photoinjector prototype for the ERL project being developed at Cornell University is foreseen to provide an average current of up to 100 mA \cite{erlgen}. Such a photoinjector requires photocathode materials with quantum yields from a few to 10\% when illuminated by visible light, so as to reduce the laser power required to a reasonable range, e.g. 10's of watts average power. Furthermore, the photocathode should have both small thermal (intrinsic) emittance, and sub-picosecond response time to produce maximally bright beams. Finally, photocathodes must be able to withstand operational damage, whether due to residual gas ion back bombardment, or high voltage breakdown events, which can both cause sputtering or disordering of the photocathode crystal.

The Boeing normal conducting RF gun operated with a K$_2$CsSb cathode, which has a typical quantum efficiency (QE) of 8\% at 527 nm, demonstrated operation at 32 mA average current, thereby setting the average current world record for photoinjectors. However, the lifetime of the photocathode was limited to a few hours due to the poor vacuum conditions of the gun \cite{dowell}. Recent measurements on alkali antimonide photocathodes show that both the thermal emittance and response time characteristics are suitable for generating high brightness beams ~\cite{therm}.

We report here the operational experience of running the Cornell ERL photoinjector at 20 mA for 8 hours, using a bi-alkali photocathode. Details on the ``\textit{post-mortem}'' analyses (SEM, AFM, X-ray fluorescence, XRD) will be given, which demonstrate that a single source of damage (e.g. ion back-bombardment) is unlikely to explain the complex of surface features. In general, we find that though this photocathode material to be more operationally rugged than GaAs activated to negative affinity, operational damage still represents an issue that needs to be addressed for delivering higher average current from photoinjectors. 

\section{Photocathode Growth and Characterization}
The bi-alkali photocathode is grown on a highly p-doped Si(100) substrate, and the UHV chamber used for deposition has been described elsewhere \cite{uhv}. Throughout deposition and use, the substrate is attached to a stainless steel puck cylinder for use in translation and registration in the gun electrodes. Our procedure for growing K$_2$CsSb photocathodes is the following:

\begin{enumerate}
\item The substrate is HF rinsed to remove the native oxide layers and subsequently heated to 600 $^\circ$C to remove the residual hydrogen passivation.
\item The temperature is lowered to approximately 100 $^\circ$C, and 10 nm of Sb is evaporatively deposited.
\item Evaporation of the K is carried out while the substrate is slowly cooling. The QE is continually measured, and K deposition ceases when a QE peak is reached.
\item When the temperature falls below 50 $^\circ$C, Cs evaporation begins until the QE again reaches a maximum. 
\item The substrate is allowed to cool down to room temperature. 
\end{enumerate}

\begin{figure*}[ht]
\includegraphics[width=1.0\linewidth,clip]{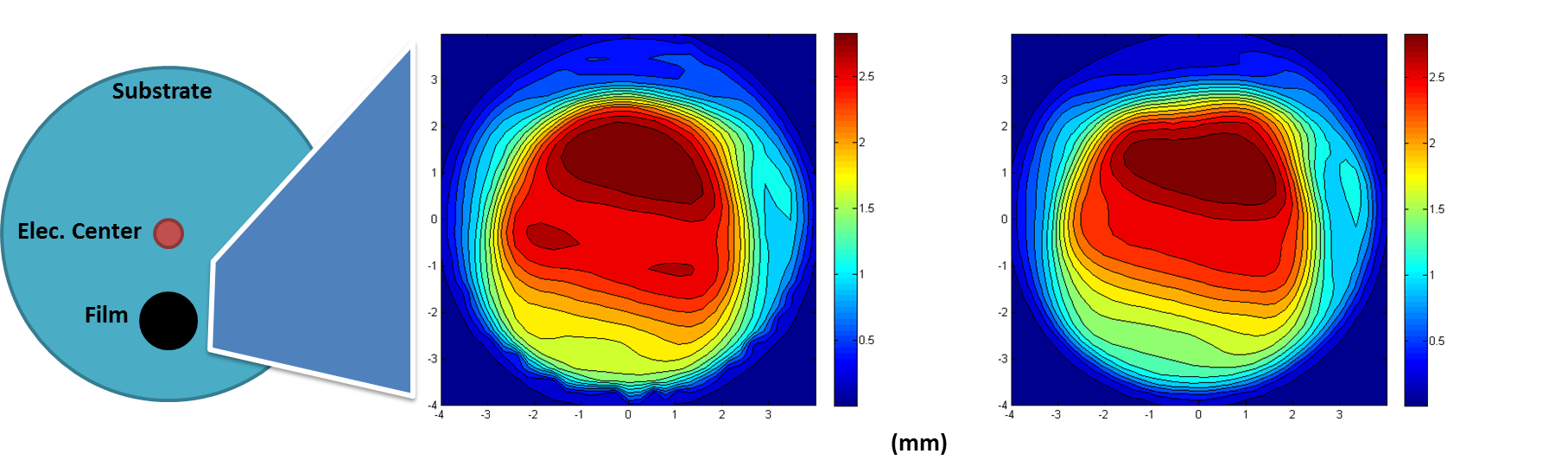}
\caption{\label{qemap} A schematic of the photocathode substrate and film is shown at left. The QE (\%) map of the photocathode is shown before running (middle) and after (right). }
\end{figure*}

One of the samples, having a diameter of about 8 mm, was grown approximately 6 mm off-center with respect to the puck symmetry axis, done in anticipation of the damage to the electrostatic center. After the growth, the puck and photocathode were transferred by means of a dedicated UHV suitcase to the load lock chamber of the ERL prototype DC gun \cite{gun}. The QE map of the photocathode was measured with a small DC laser at 532 nm once in the load lock chamber, and before insertion into the gun, and is shown in Fig. (\ref{qemap}).

\begin{figure*}[ht]
\includegraphics[width=1.0\linewidth,clip]{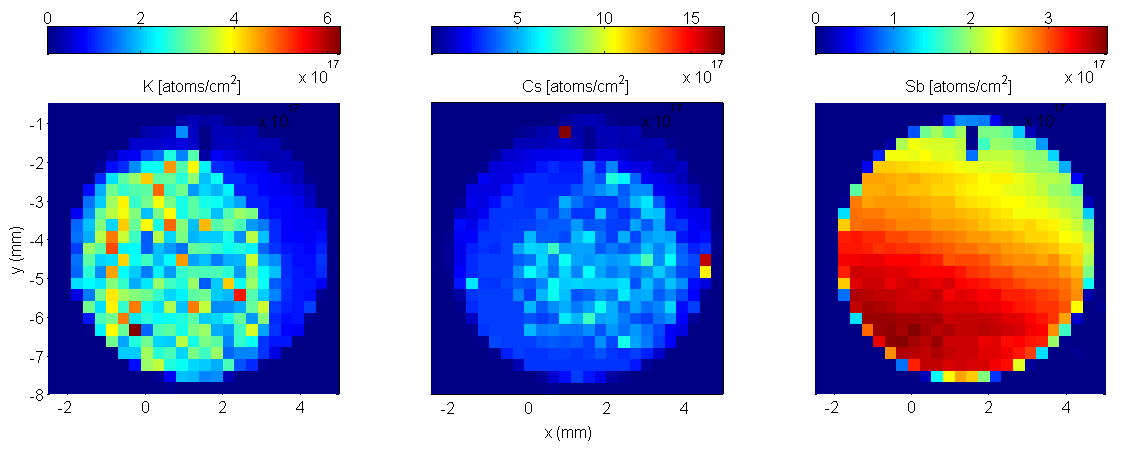}
\caption{\label{xrfmaps} XRF maps for K (left), Cs (middle), and Sb (right). A notch had been scribed in the sample for orientation purposes.  The stoichiometry, normalized to the Sb areal density, averaged over the active area was found to be K$_{0.8}$Cs$_{1.29}$Sb, where the values for the potassium areal density should be treated as a lower bound on the actual value (see text).}
\end{figure*}

The QE peak (3\%) as well as the QE uniformity were deemed non-ideal, likely due to misalignment of the evaporating sources within the deposition chamber. The poor uniformity was verified with XRF measurements performed at the F3 line at the Cornell High Energy Synchrotron Source (CHESS). The F3 station used a 10 keV X-ray beam with approximately 10 $\mu$m diameter, incident upon the sample surface with a coincidence angle of 45 degrees. An energy dispersive Vortex ME4 silicon drift detector fabricated by SII Nanotechnology
measured X-ray fluorescence from the sample, which is generally capable detecting trace elements down to the parts per million level. The sample was scanned with the use of a 2D translation stage, yielding 2D maps of the atomic areal density, and thus ultimately the chemical composition. In order to calibrate the fluorescent flux incident on the detector to yield the atomic areal density, we prepared thin film test samples on Si substrates for Sb, K, and Cs. Rutherford backscattering measurements were then used to determine the areal density of the thin film test samples, and these test samples were then run under XRF prior to analyzing the actual photocathode sample.  The XRF maps are plotted in Fig. (\ref{xrfmaps}). Several causes contribute to a large systematic uncertainty is inherent in these measurements. First, the RBS beam scatters off a small area of the substrate, and the test films analyzed for calibration also include the large variations in surface concentration. Second, these measurements were performed in air after the cathode's use in the injector, and it is unknown how interaction with air will affect the areal density of the film constituents. Finally, in the case of the potassium test film, it was seen that the potassium RBS scattering peak lies too close to that of Si to allow a precise determination of the areal density. Thus, the values quoted for potassium should be viewed only as lower bounds; it is possible that the stoichiometry is closer to the expected value of K$_2$CsSb.

The atomic areal density maps clearly show a non-uniform distribution of the metallic species, implying some misalignment of the substrate with the evaporative sources, and thereby creating a nonuniform QE distribution. Nevertheless, this photocathode had sufficiently high QE to be operated as a first test for delivering mA level current, and to investigate its ruggedness with respect to long operation times. It is interesting to note that though clearly visible to the naked eye (see below), the features of damage from use are not evident in the XRF maps.

\section{High Current Run Performance}

\begin{figure*}[ht]
\includegraphics[width=1.0\linewidth,clip]{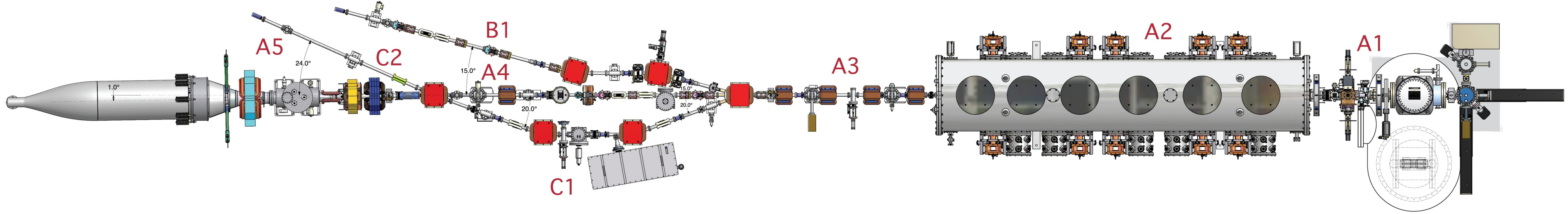}
\caption{\label{inj} A schematic of the Cornell ERL photoinjector. Beam travels from right to left. }
\end{figure*}

\begin{figure*}[ht]
\includegraphics[width=1.0\linewidth,clip]{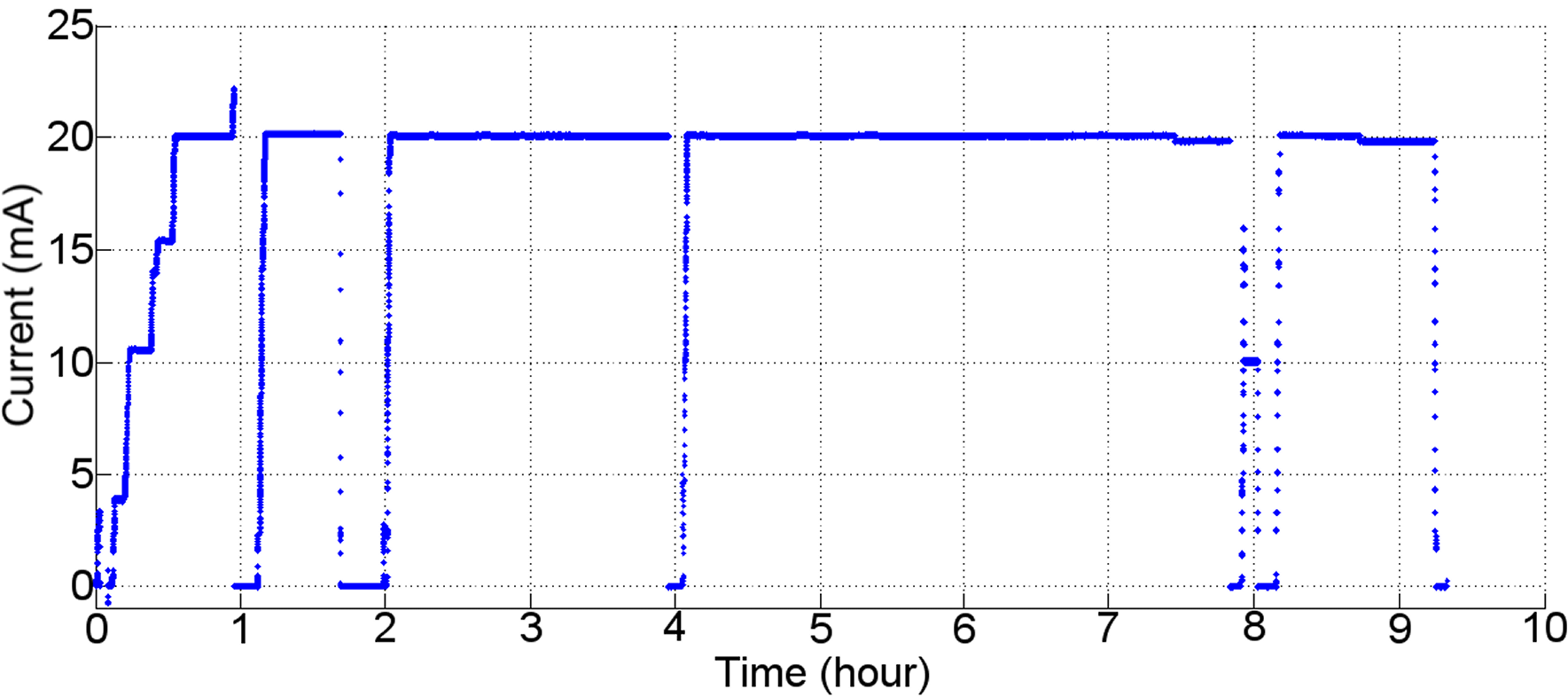}
\caption{\label{cur} Current as a function of time for the run in question.}
\end{figure*}

A schematic of the Cornell ERL photoinjector is shown in Fig. (\ref{inj}), and is described in \cite{proto}. The photocathode is inserted directly into the electrostatic center of the Pierce electrode structure of a DC gun operating at 250 kV, and the photocathode is illuminated by a  1.3 GHz repetition rate laser operating at 520 nm. The 250 keV electrons pass through two emittance compensation solenoids and a normal-conducting buncher cavity before entering the 5 two-cell SRF cavity cryomodule, which boosts the energy to about 5 MeV. The average current as a function of time for this run is plotted in Fig. (\ref{cur}).

The current during the run was limited to 20 mA, though the combination of deliverable laser power ($>$10 W) and 2.5\% QE allowed for higher currents (20 mA requires about 2 W average laser power at 520 nm). The current was held constant by a feedback loop which varies the laser power on the cathode via a Pockels cell, to account for variations in upstream laser power, as well as gradual degradation of cathode quantum efficiency. As visible in Fig. (\ref{cur}), during injector operation, a number of vacuum burst events triggered the fast shutdown safety system. After each trip, the beam current was successfully restored to 20 mA.

After an approximate run time of 8 hours, and a total charge delivered of about 575 Coulombs, the photocathode puck was retracted into the UHV preparation chamber for visual inspection and QE measurement. It should be noted that the QE, as determined by the delivered laser power (data not available), showed no apparent degradation and did not limit the length of the run; the run ended at the preference of the operators. The post-run QE map in Fig.(\ref{qemap}, right) shows a small, nonuniform decrease, with more significant damage toward the gun electrostatic center, which in the above maps is upward.

\section{Post-run photocathode characterization}
\subsection{Visual inspection}
The most salient feature of the photocathode appearance upon removal from the gun was the presence of a highly damaged area approximately at the electrostatic center of the gun. A photograph taken in air of the sample after removal from the gun is found in Fig. (\ref{pic}). Naked eye inspection of the surface reveals different regions with well defined characteristics:

\begin{figure}[t]
\includegraphics[width=0.8\linewidth,clip]{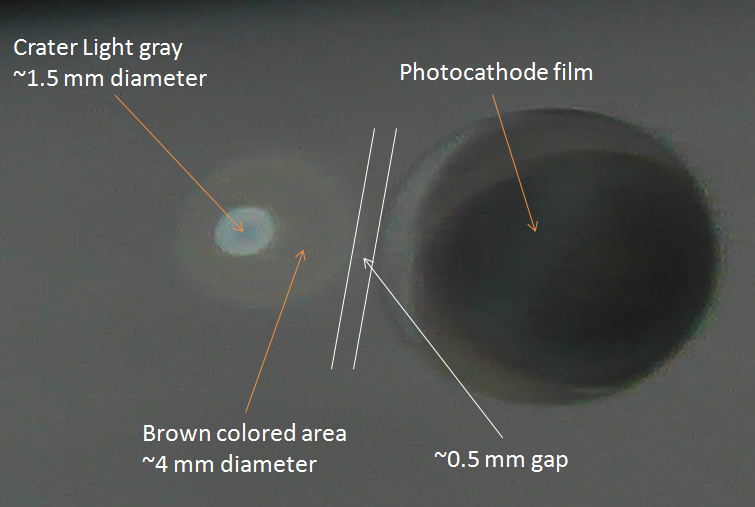}
\caption{\label{pic} A close-up photo of the photocathode film and substrate after use. }
\end{figure}

\begin{itemize}
\item \textit{The Photocathode Film:} The three non-concentric rings of the bi-alkali film indicate the poor alignment of the substrate with the evaporative alkali sources. This, along with the XRF species maps, supports the hypothesis that the poor QE performance is due in part to poor alignment.

\item \textit{The Central Damage:} Located roughly at the electrostatic center of the puck, approximately 7 mm from the center of the photocathode film, this region has a well defined circular shape. Naked eye inspection suggests extreme surface roughening due to the loss of the typical mirror-finish of the polished Si substrate.

\item \textit{The Brownish Area:} This area asymmetrically surrounds the central damage area, and presents a brownish color and loss of mirror-finish. The centroid of this region is shifted towards the site of electron emission.
\end{itemize} 

\subsection{Profilometry, AFM and SEM Results}

Atomic force microscopy (AFM) was carried out using a Veeco Dimension 3100 instrument in tapping mode. Scanning electron microscopy  was performed with a Cambridge Stereoscan 200 microscope. Profilometry was performed using a Stylus Surface Profiler Tencor Alpha Step. 

\begin{figure}[t]
\includegraphics[width=1.0\linewidth,clip]{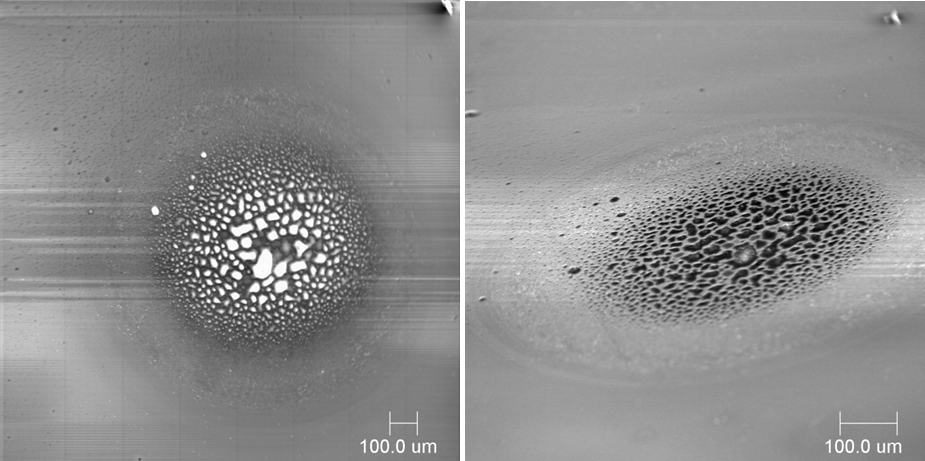}
\caption{\label{semdamage} SEM micrographs of the central damage area at different substrate inclination angles. }
\end{figure}

SEM micrographs of the central damage area are shown in Fig.(\ref{semdamage}). According to previously reported work on GaAs and GaAsP we believe that the central damage area has been caused, at least in part, by the back-bombardment of residual gas ionized by the beam \cite{grames, othergrames}. The ions in question are likely created between the photocathode and the entrance to the first SRF cavity, and upon entering the gun's accelerating field, can impact the surface with up to 250 keV of energy.

Furthermore, given that the central damage is predominantly at the electrostatic center, it is likely that the majority of the ions are produced once the beam has been aligned on its symmetry axis, which is done with corrector magnets after the exit of the gun. It is of interest to note that both the height and size of the microstructures increase as you approach the electrostatic center of the sample. This distribution of microstructures should be due to some nontrivial combination of the ion beam transverse profile with the integrated ion dose.

\begin{figure}[h]
\includegraphics[width=0.5\linewidth,clip]{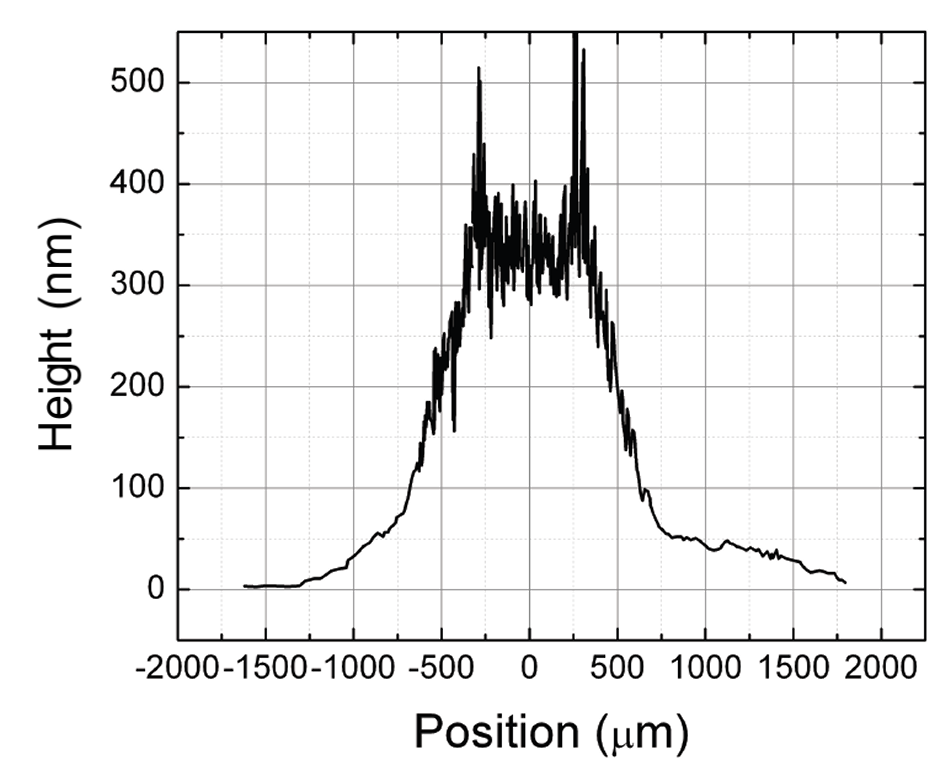}
\caption{\label{prof} Surface profilometry of the central damage area, indicates the presence of significant surface roughening. Due to the large tip size, the profile data should be interpreted as sampling the local maxima of the surface profile.}
\end{figure}

The surface profile from contact profilometry is shown in Fig.(\ref{prof}). While intuition suggested that the central damage area was a crater due to ion sputtering of material,  the surface profilometry suggests that the damage site is raised (or simply roughened) with respect to the ordered substrate crystal. It is likely that the thickness of the tip obscures the fast oscillations of the surface roughening, and thus the average height of 350 nm should be interpreted as an average of surface maxima. 

\begin{figure}[ht]
\includegraphics[width=1.0\linewidth,clip]{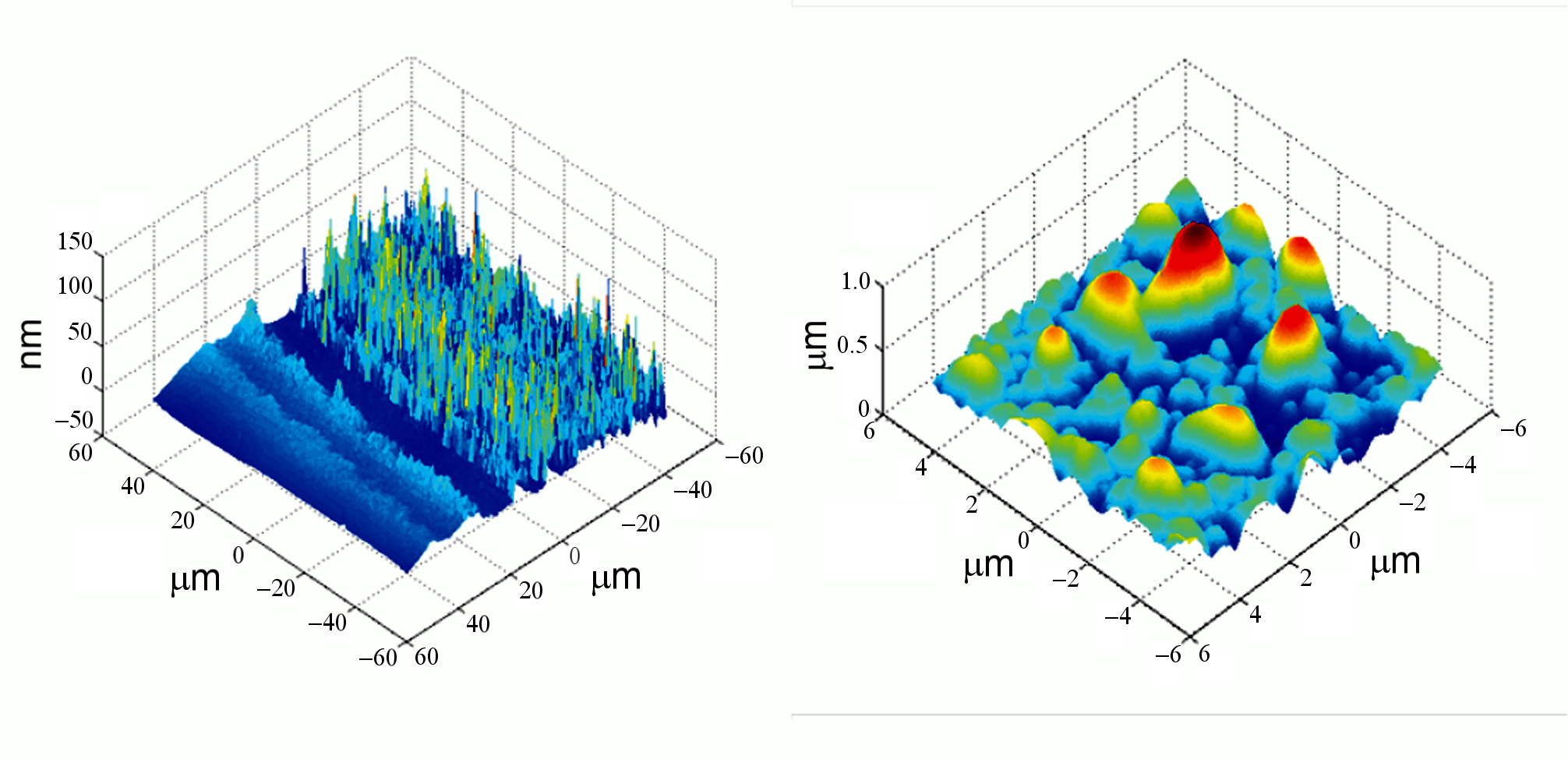}
\caption{\label{afm} Atomic force micrographs of the central damage area. Left: 100$\times$100 $\mu$m$^2$ scan at the interface of the brownish area and the central damage site. Right:  10$\times$10 $\mu$m$^2$ scan at the mean distance between the center and edge of the central damage site. }
\end{figure}

Atomic force microscopy was carried out over an area of 100$\times$100 $\mu$m$^2$ near the edge of the central damage, see Fig.(\ref{afm}, left). In these images the fast oscillations of the surface roughening are apparent. Moving from the border to the central part of the damage, the surface becomes more and more rough, but the resolution of this scan was unable to reveal details of the surface in the central part of the damage.  Thus, we also performed surface scans with a smaller resolution over an area of 10$\times$10 $\mu$m$^2$ in an intermediate position between the edge and center of the central damage site, shown in Fig.(\ref{afm}, right). From this scan, we see that the surface exhibits no sharp point-like peaks, and has an rms surface roughness of about 100 nm.

\subsection{X-ray Diffraction Results}

Structural defects and damage induced in the Si crystal due to running conditions were studied using X-ray diffraction at the C1 line of the CHESS facility. The C1 line uses a monochromatic 15 keV highly parallel x-ray beam with a transverse size of 1$\times$1 cm$^2$, directed on the Si(100) surface. The Si wafer was held on a 4 circle goniometer for precise alignment. The diffracted X-rays were collected using a CCD camera, allowing reconstruction of the 2D diffraction maps all along the illuminated area. The angular resolution for the diffraction angle used during these experiments was  $5 \times 10^{-4}$ degree. The sample was aligned so as to view the diffraction of the Si(004) planes.

\begin{figure}[ht]
\includegraphics[width=1.0\linewidth,clip]{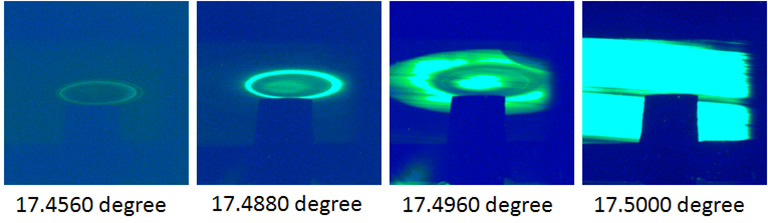}
\caption{\label{xrdsnaps} False color photographs from the XRD CCD camera at various diffraction angles, illustrating the variation in lattice spacing as a function of distance from the central damage spot. }
\end{figure}

\begin{figure}[ht]
\includegraphics[width=0.5\linewidth, clip]{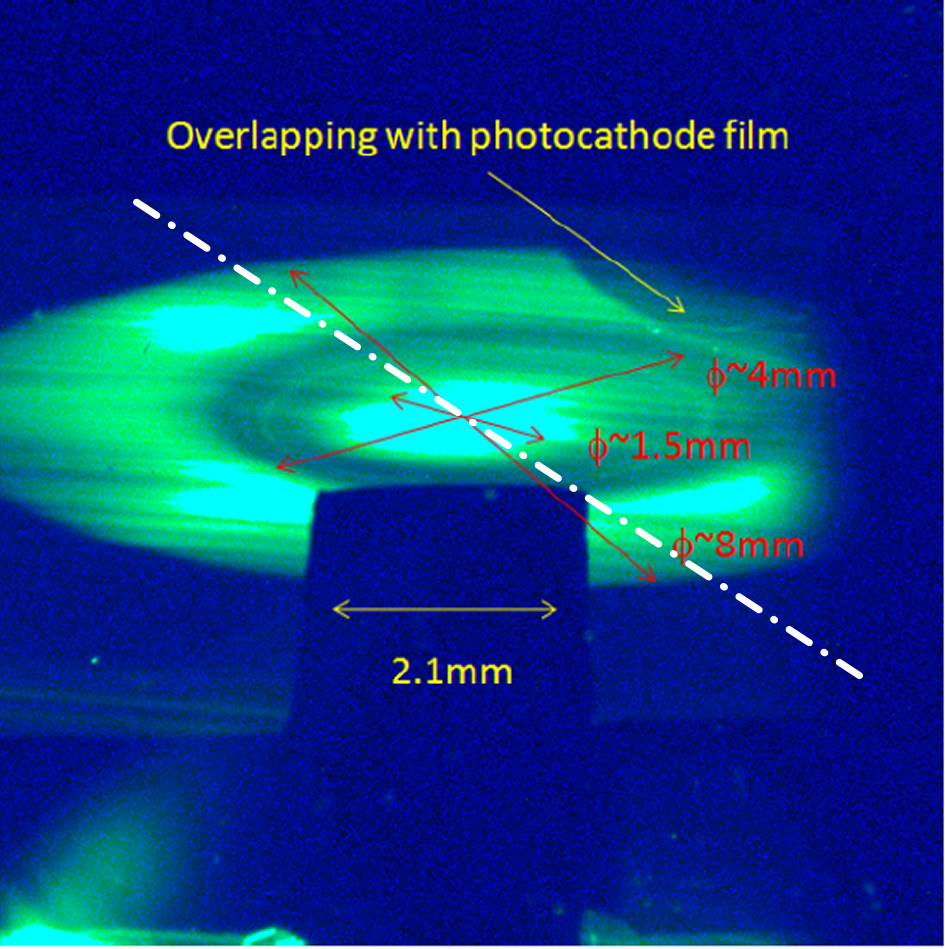}
\caption{\label{xrdlen} Various diameters of the regions of interest from the XRD CCD snapshots. The white dotted line is the diameter over which data was sampled for Fig. (\ref{xrdtot}).}
\end{figure}

Fig.(\ref{xrdsnaps}) shows a false color image taken with the XRD CCD at various substrate inclination angles, indicating a nearly radial distribution of lattice spacing, measured from the center of the damage site. As the diffraction angle, $\theta$, approaches that of the Si(004) diffraction peak (see $\theta=17.4960^\circ$ ), a round area with well defined edges emerges in the upper right corner of the image, which was the location of the bi-alkali film, indicating lattice damage overlaps with the active area of the cathode.  A rectangular lead stripe of thickness 2.1 mm was overlaid on the sample for length calibration; Fig.(\ref{xrdlen}) reports the approximate diameters for the different areas of diffraction peaks.

\begin{figure}[ht]
\includegraphics[width=0.8\linewidth, clip]{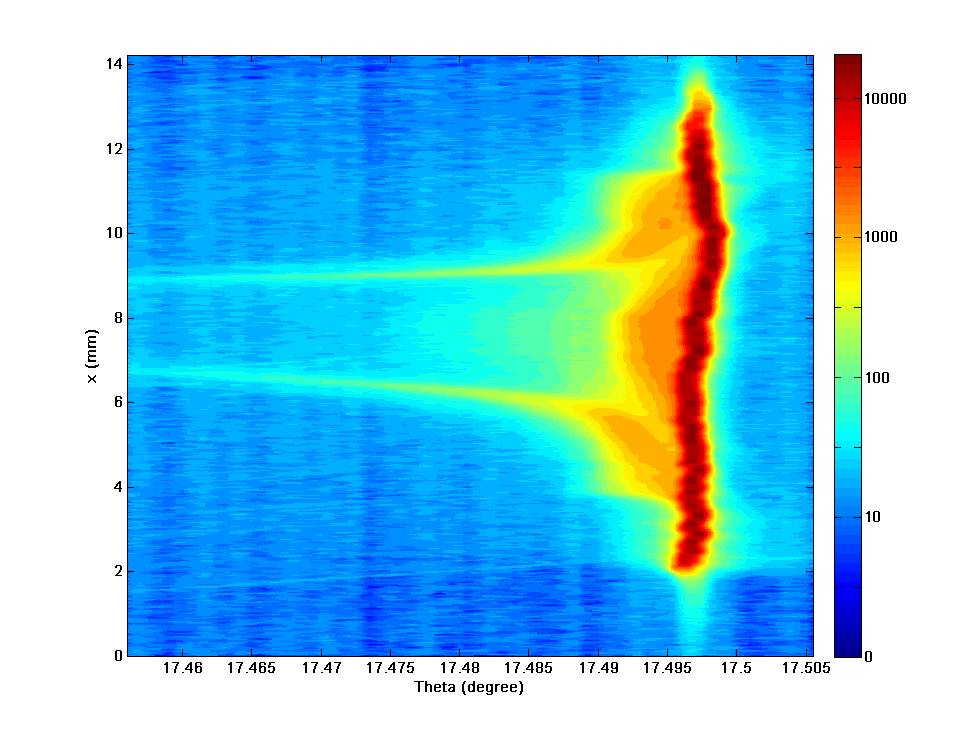}
\caption{\label{xrdtot} Counts on the CCD detector as a function of the position along a diameter of the circular damage region ($x$, vertical axis) and the diffraction angle. }
\end{figure}

Given that the distribution of crystal strain regions is radial, we can collapse the spatial data down to one dimension, and plot the number of counts on the CCD detector as a function of both position along the diameter of the damage region and diffraction angle. This is shown in Fig.(\ref{xrdtot}). The induced crystal strain is proportional to the deviation of the diffraction angle from the Si(004) peak (rightmost red region). The crystal strain is symmetric about the damage site center,  and is sharply peaked on either side of the electrostatic center (here at $x\sim8$ mm).

The smallest diffraction angle corresponds to the region of largest atomic spacing. The annular region that first diffracts (at left in Fig. \ref{xrdsnaps}) appears to be the region of largest damage. It appears unlikely from a beam dynamics perspective that the backstreaming ion distribution should be annular (see section below). Thus, the surface region inside the annulus may be the site of complete amorphization of the surface, and thus not subject to diffraction. This region (as well as the region beyond the initial annulus) does begin to difract at slightly larger angles, which may be due to the penetration of the x-rays below the amorphous layer, to a layer beneath which is strained but still crystalline. This is supported by the fact that at the Si(004) diffraction peak, the entire substrate including damage center diffracts, which may be due to the bulk, undamaged structure beneath (on the scale of $\mu$m) the damaged surface. Given that the extinction depth of the incident x-rays is on the order of 10 $\mu$m (which also depends of atomic spacing), it is important to note that the diffraction vs. depth is unavailable from these scans. 

\section{Estimation of effects of Ion Back-Bombardment}

It is likely that the impact of back-propagating ions along the beamline is at least partially responsible for the damage features described above. To determine the amount of ions present in the beamline during high current operation, we use the partial pressures of different chemical species as measured by a residual gas analyzer (RGA) within the DC gun chamber, the most prominent of which are shown in the Table (\ref{pptable}) below. It should be noted that the RGA measurement did not take place during the high current run; this is a source of uncertainty, but we believe the values to be representative at least to order of magnitude. Furthermore, with known gas conductances, pumping speeds, and outgassing rates, we can estimate the pressure profile along the beamline, and we assume that the partial pressures will scale linearly with the overall pressure.	To calculate the total number of ions, we then use the Bethe ionization cross section for each species, which yields approximately $1.85\times10^9$ ions/Coulomb, and a total ion dose of $1.1\times10^{12}$ over the course of the run. We assume these ions to have an initial velocity that is thermal, and to impact the surface of the photocathode with the full gun voltage (250 keV). It should be noted that this should be treated as extreme lower bound on the possible number of ions present during the run. First, this estimate was based on the RGA spectrum of the gun chamber, which houses a number of non-evaporative getter (NEG) pumping modules, as well as conservative outgassing rates; the partial pressures of residual gas after the gun may be orders of magnitude larger. Secondly, this estimate does not take into account the number of vacuum trip events throughout the run, which may also affect the number of ions on the order of magnitude level.

The 250 keV gun energy is within the range of typical ion implantation values used for doping. A recent overview paper of the ion implantation induced damage,  amorphization, and recrystallization in Si illustrates several effects that we believe may be observed in the damage site \cite{overview}. To understand the effect of each ion species independently, we ran several simulations using the TRansport of Ions in Material (TRIM) software package \cite{srim}. As TRIM does not support the use of molecular ions, we made the simplifying assumption that all molecules will be fully chemically cracked upon impact with the photocathode surface. Under this assumption, the composition of the ion beam is given at the bottom of Table (\ref{pptable}). Simulations were performed with 250 keV carbon, hydrogen, and oxygen ions. A final run of 250 keV singly ionized silicon was performed to allow comparison with the implantation literature.

\begin{table}
\caption{Top: Partial pressures of the dominant residual gas species, as measured by RGA within the DC gun chamber. Middle: Ion beam composition, calculated from partial pressure and Bethe ionization cross section. Bottom: Percent composition of a fully cracked ion beam.}
  \begin{tabular}{ |c | c |  }
 \hline

Species & Partial Pressure (torr)\\ \hline
H$_2$ & $1\times10^{-11}$  \\
H$_2$O & $1\times10^{-12}$ \\ 
CO & $5\times10^{-13}$  \\
CO$_2$ & $1\times10^{-13}$ \\ \hline \hline

Species & Percent composition \\ \hline
H$_2^+$ & 75\% \\
H$_2$O$^+$  & 5\% \\
CO$^+$ & 15\% \\
CO$_2^+$ & 5\%    \\ \hline \hline

Cracked Species & Percent composition \\ \hline
H$^+$ & 64\% \\
O$^+$ & 21\% \\
C$^+$ & 15\% \\
\hline
\end{tabular}
\label{pptable}
\end{table}

\begin{figure}[ht]
\includegraphics[width=0.5\linewidth, clip]{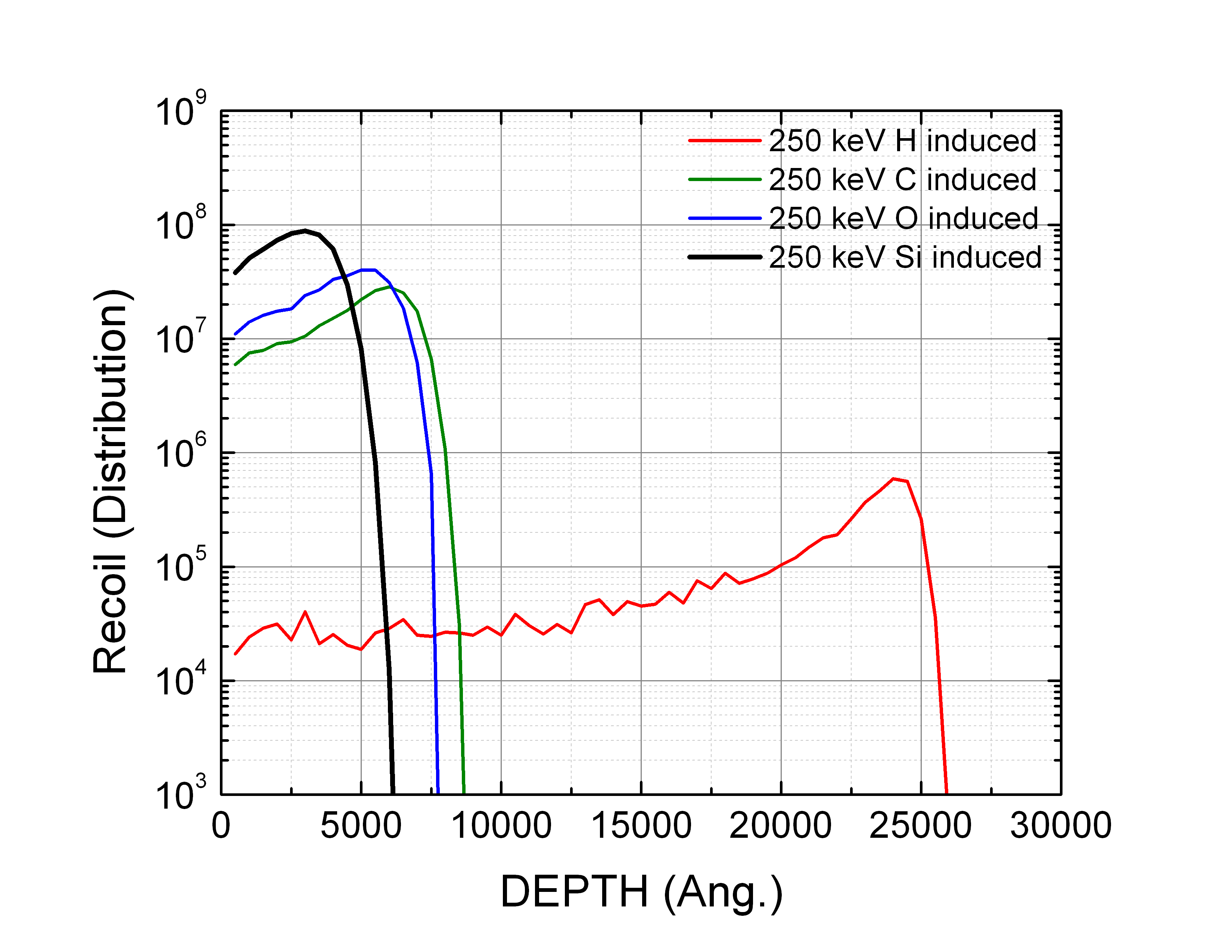}
\caption{\label{trim} The number of atomic recoils as a function of penetration depth for a variety of implanted ions. }
\end{figure}

The results of the TRIM simulations were performed in \textit{Surface Sputtering and Monolayer Collision} mode, and are shown in Fig.(\ref{trim}). It shows the number of atomic recoils for each species as a function of the ion penetration. It is clear from the TRIM data that the effect of hydrogen ions on the surface film is likely to be very small in comparison to that of carbon or oxygen, given both the small number of recoils and the relatively large penetration depth. Also, we conclude that we are well justified in applying results from Si self implantation literature, as the integrated number of atomic recoils of both carbon, oxygen, and hydrogen are comparable to that of Si. TRIM also reports the sputtering yield from the various implantation species; hydrogen was found to have a negligible sputtering yield, whereas carbon and oxygen ions at 250 keV have 0.16 and 0.18 sputtering yield, respectively, which alone appear to be inadequate to describe the features of the damage area.

\begin{figure}[ht]
\includegraphics[width=1.0\linewidth, clip]{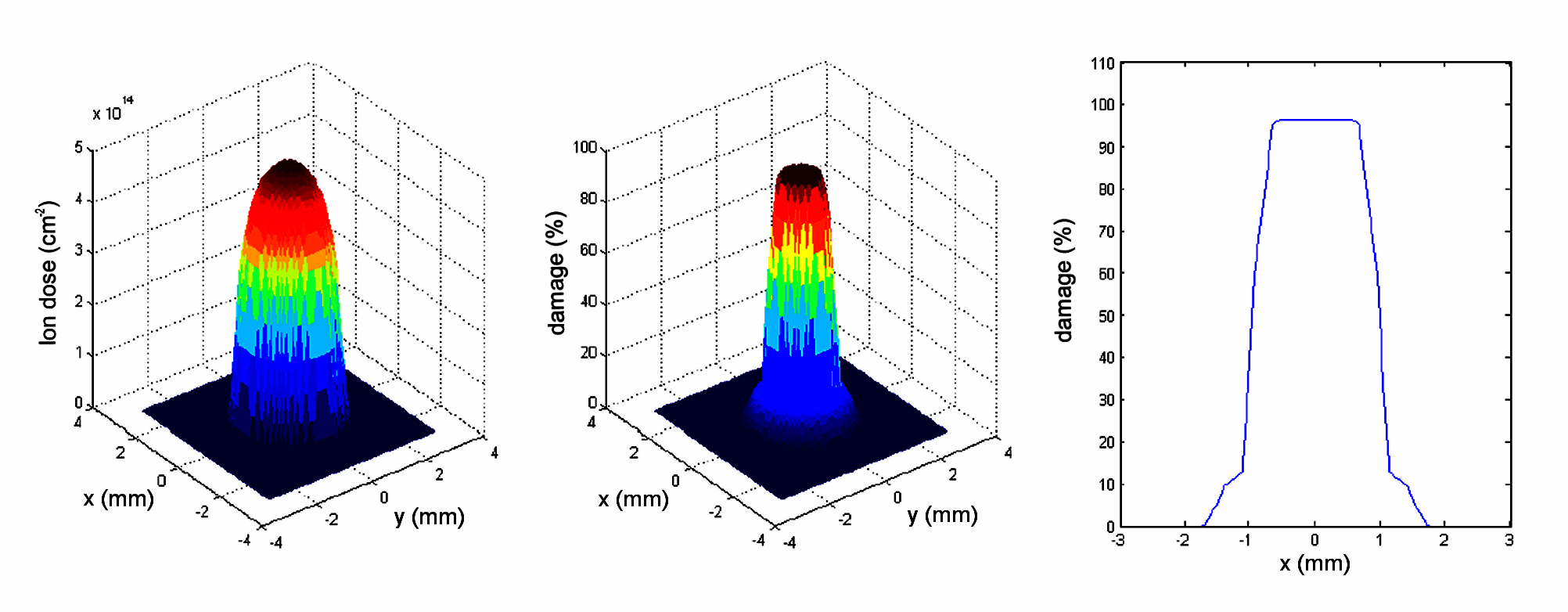}
\caption{\label{ions} Left: The 2D dose distribution of ions, calculated from the simulated electron beam profile at the DC gun anode. Middle: The damage profile, or percent amorphization of the Si using the ion beam profile and a superlinear model. Right: a cross section of the damage profile.   }
\end{figure}

In order to understand the spatial distribution of damage due to ion back bombardment, we must determine the electron beam profile in the locations where cathode-sensitive ions are likely to be produced. To do this, we used ASTRA \cite{astra} to calculate both the beam envelopes and transverse distributions for the parameters of the run, including a realistic transverse and temporal laser profile. We assume for simplicity that the electron beam transverse profile at the anode is the ion beam distribution that is incident on the photocathode. This beam profile, along with the partial pressure and ionization cross section, yields the 2D ion dose distribution, which is plotted in Fig.(\ref{ions}, left). 

According to \cite{multiple}, if we define damage as the amount of RBS dechanneling measured with a partially amorphous sample, the damage profile (on a scale from 0-1, or completely dechannelled) can be calculated directly from the ion dose. It is important to note from \cite{multiple} that the damage does not scale linearly with the dose. For doses below a critical value (for Si ions at 230 keV, doses below $4\times10^{14}$ ions/cm$^2$), the damage grows slowly and linearly with increasing ion flux. Up to this point, it is thought that the crystalline Si (c-Si)  develops small regions of amorphous silicon (a-Si) by ion impact, but their density is so low that a partial recombination of point defects alleviates damage growth. However, beyond the critical value, the ion damage slope increases sharply, into the so-called ``superlinear" regime, in which minimal recombination occurs. As the ion dose increases, the damage slope saturates as it approaches unity, or complete dechanneling.

Using this information, a threshold dosage of $4\times10^{14}$ ions/cm$^2$, and the same damage vs. dose slope as in \cite{multiple}, we calculate the damage profile shown in Fig.(\ref{ions}, middle, right). The sharp increase in the damage is due to the superlinear response of the amorphization. We should note that in order to reach the onset of the superlinear damage, we used a number of ions only 10 times larger than the lower bound estimate, according to the rationale above. Upon inspection of the XRD data in Fig.(\ref{xrdtot}), the sharp decrease in the number of counts for large angular offsets from the Si(004) peak near the center of the damage site may be due to the complete amorphization of the Si substrate surface, thereby suppressing the diffraction in that region, due to the onset of the superlinear damage regime. Thus, there is good qualitative agreement between the ion back-bombardment calculated damage profile and both the profilometric and XRD maps.

\section{Remaining Questions}
\begin{figure}[ht]
\includegraphics[width=0.5\linewidth, clip]{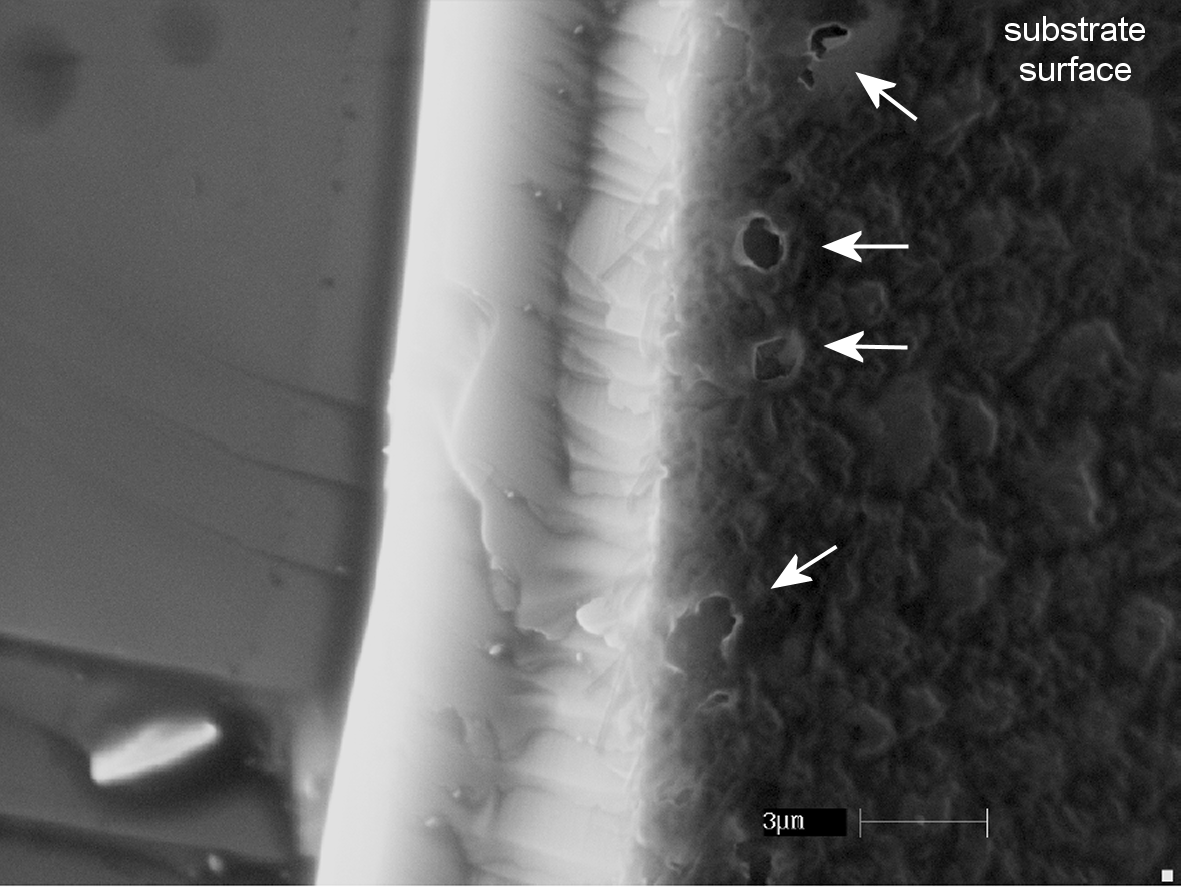}
\caption{\label{sem1} SEM micrograph of the substrate cross section, beneath the damage site. The surface of the substrate is towards the right. Pore structures are indicated by arrows.  }
\end{figure}

\begin{figure}[ht]
\includegraphics[width=0.5\linewidth, clip]{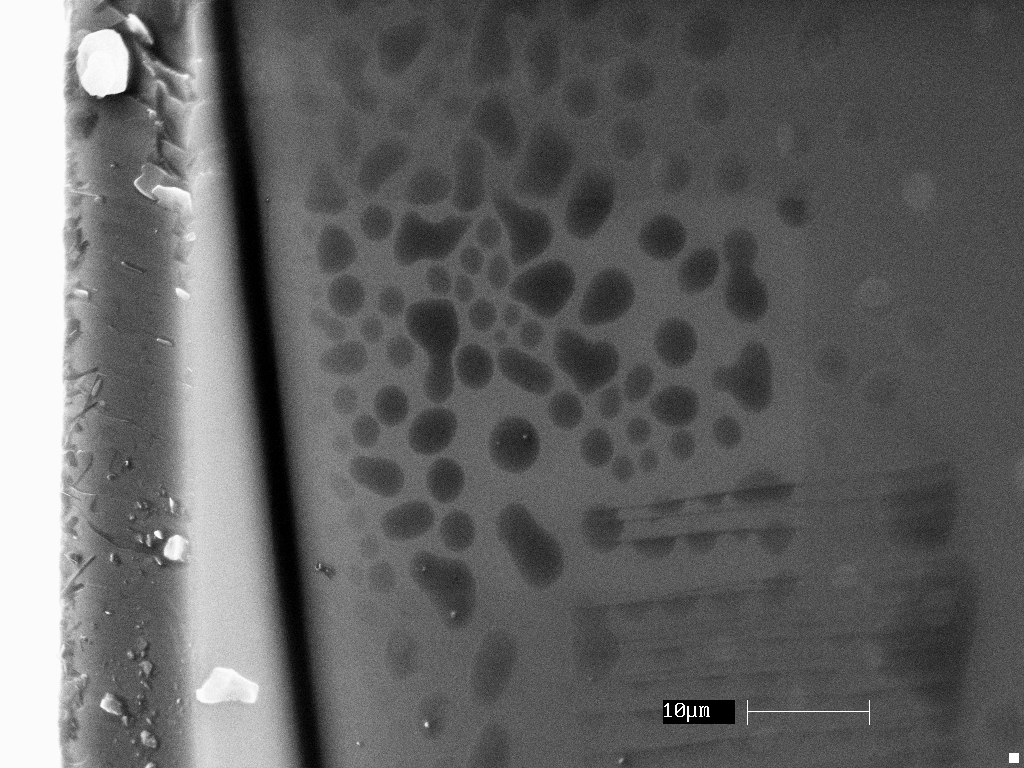}
\caption{\label{sem2} A lower resolution, larger scan range SEM micrograph beneath the damage site. The surface of the substrate is towards the right.  }
\end{figure}

In order to investigate possible means of the formation of the surface roughening/protrusions, the sample was hand-cleaved through the center of the damage area, and a high-resolution SEM micrograph was taken of the substrate cross section near the surface. These micrographs are shown in Fig. (\ref{sem1}) and Fig. (\ref{sem2}). Fig. (\ref{sem1}) indicates the presence of pore structures below the Si substrate. Similar pore production has been reported in literature as a consequence of ion implantation with doses several orders of magnitude larger than the ones we calculate incident upon our photocathode \cite{pores}, which suggests that our estimate of the total ion dose may be far lower than the actual value.  Furthermore, Fig. (\ref{sem2}) indicates the presence of bubble-like regions 10's of $\mu$m deep (deeper than the expected penetration of 250 keV ions), which are present only below the damage region, and not below hand-cleaved pristine Si. The nature of these structures is still currently under investigation.

\section{Conclusions and Outlook}
In this paper, we have described the successful operation of the Cornell ERL injector prototype at 20 mA for an 8 hour period, utilizing a K$_2$CsSb photocathode. A successful recipe for photocathode growth is described, and it was seen that the QE had little to no degradation as a result of the high current operation. The photocathode substrate itself exhibited striking damage features at and near the electrostatic center, which were characterized by AFM, XRF, XRD, SEM and contact profilometry. We determined that the central most damage area was not in fact a sputtering crater, but surface roughening which protrudes above the substrate, which is corroborated by ion back bombardment TRIM simulations, which suggest that material sputtering from ions is minimal. We found that the substrate damage was both a combination of crystal strain and amorphization, at least in part from ion back bombardment. This was justified by the use of a simple simulation of the damage profile based on the electron beam transverse distribution, which agreed well with the XRD and profilometry damage measurements. Several open questions remain, including the mechanism of formation of the damage protrusions, as well as several anomalous structures detected below the damage site surface. In general, we find that not only is the mitigation of operational damage of photocathodes in photoinjectors crucial to the production of high average brightness beams, but also that the nature of the photocathode damage is more complex than has been previously characterized. 

Multiple avenues exist for future study. The use of metallic substrates, such as molybdenum, is currently under investigation at Cornell, and may prove to be more resilient to the operational damage. However, it should be noted that the perfect crystallinity of the Si substrates enabled the use of diffraction imaging, which provided valuable damage information. Such information should also be available from the negative affinity GaAs photocathodes used in a number of laboratories currently. The effect of ion back bombardment may be better characterized by a dedicated experiment, in which a known quantity of pure control gas (such as Ar, CO$_2$, CH$_4$, CO, etc) is introduced in the beamline during high current running. Such an experiment would be best performed with the ability to resolve the quantum efficiency of the cathode as function of both time a position on the substrate, and would likely require the extraction of significant charge. A dedicated beam plan for such a high current study is currently under development for the Cornell ERL. 

\section{Acknowledgments}
The ERL photoinjector program at Cornell University is supported by the NSF, under award DMR-0807731. This work is based upon research conducted at the Cornell High Energy Synchrotron Source (CHESS) which is supported by the National Science Foundation and the National Institutes of Health/National Institute of General Medical Sciences under NSF award DMR-0936384.

\bibliography{20marunfinal}{}
\bibliographystyle{apsrev}

\end{document}